\documentclass[floatfix,preprintnumbers,amsmath,amssymb]{revtex4}

\usepackage{graphicx}
\usepackage{dcolumn}
\usepackage{bm}
\begin{document}
\title{Colour Confinement and Deformed Baryons in Quantum Chromodynamics}
\author{Syed Afsar Abbas}
\affiliation{Department of Physics, Aligarh Muslim University, Aligarh - 202 002, India.\\
email:drafsarabbas@yahoo.in}
\begin{abstract}
The confinement of coloured entities in Quantum Chromodynamics (QCD) is traced to colour singletness of the observed entities. This is believed to arise from colour singlet state of quark-antiquark for mesons and a fully colour antisymmetric state for baryons. This demands a spherically symmetric baryon in the ground state. However it is pointed out that a deformed baryon in the ground state has been found to be extremely successful phenomenology. There are convincing experimental supports for a deformed nucleon as well. This means that something has been missed in the fundamental theory. In this paper this problem is traced to a new colour singlet state for baryons which has been missed hitherto and incorporation of which provides a consistent justification of a deformed baryon in the ground state. Interestingly this new colour singlet state is global in nature.
\end{abstract}
\maketitle 
Quantum Chromodynamics (QCD) has been an extremely successful 
theory of the strong interaction. However colour confinement
still remains an unsolved problem of QCD. In spite of intense
efforts one has not been able to understand it fully within QCD.
Because of this it has even been compared to Fermat's Last Problem
in mathematics\cite{a}.

If not understood fundamentally, phenomenologically one has attained
a reasonable understanding of confinement in QCD\cite{b}. As per our 
best understanding, colour singletness is the basis for confinement.
Mesons are confined due to colour singlet states for the
quark-antiquark. For three quarks one obtains saturation in colour space
by having the colours of the quarks constituting a totally antisymmetric wave function.
Hence for baryons, the wave function in the rest of the degrees of 
freedom is fully symmetric in the exchange of any two quarks.

As per this framework of confinement of three quarks in a baryon, these should sit in a spherically symmetric s-state\cite{c}. This model does pretty well, say, fitting the magnetic moments of nucleon quite well etc. This model forms the basis of the quark model~\cite{c} and may be called the canonical phenomenological model of the baryons. Note that spherical symmetricity of baryons is the backbone of this model.

Though pretty successful, it has had failures too. This has led to it being labeled as Naive Quark Model (NQM). This has prompted people to go beyond NQM and propose various other phenomenological models.

However, the "Naive" tag on quark model is quite unjustified if we are able to rectify its flaws within the quark model itself. For example, it was noted by Glashow\cite{d} that the quark model does pretty badly for the axial vector coupling constant $g_A$ where, in quark model its value is $5/3$\cite{c} while the experimental number is $1.24$. Similar problem exist for D/F parameter of the weak decay of baryons. To rectify the problem Glashow\cite{d} suggested an appropriate amount of mixture of D-state in the nucleonic ground state.
\begin{equation}
|N\rangle~=~\sqrt{1-P_D}~|N_S\rangle~+~\sqrt{P_D}~|N_D\rangle
\end{equation}
where $|N_S\rangle$ and $|N_D\rangle$ give the spherically symmetric part $(56,0^+)$ and the D-state admixture of $(70,2^+)$ to the nucleon respectively (see Appendix). The first one is the spherically symmetric part and the second one the deformed part. $P_D$ is the D-state probability. This clearly suggests that the nucleon is deformed in the ground state itself. Glashow found that a $P_D$ of $\sim 0.25$ yielded good fits to the experimentally measured $g_A$ and D/F ratios. Later Vento et al.\cite{e} found that the ratio of $\pi N\Delta$ to $\pi NN$ coupling constant is fitted very well with the same $P_D\sim0.25$.

The present author has found that the same $P_D\sim0.25$ does very well for a large number of physically relevant quantities in the quark model:\\
(a) A comprehensive fit to allowed semileptonic decays of all the $1/2^+$ baryon octet\cite{f}.
(b) Pion-Nucleon-Delta coupling constant, $E2/M1$ ratio, double delta coupling constant, constituent quark masses etc.\cite{g}.
(c) A consistent understanding of the contentious spin structure of the nucleon\cite{h}.
(d) Consistent understanding of the magnetic dipole strength and the Gamow-Teller strengths in nuclei\cite{i}.

Such comprehensive fits to a large number of physically measured quantities is very impressive. Most remarkable is the fact that it is just one parameter $P_D\sim0.25$ which does the job for all. We therefore feel that a nucleon deformed in the ground state should be taken seriously. Experimentally too there are strong evidences of deformed baryons in the ground state. For example recent improved experimental determination of the $p\rightarrow\Delta^+$ transition quadrapole moment\cite{j} gives convincing evidence of an intrinsically deformed nucleon. This is in conflict with our current understanding of confinement of baryons, it should be spherical. So it means that we are actually missing some fundamental aspect of the reality of confinement. Which aspect of confinement could we be missing which could account for such large $(25\%)$ deformation of the nucleon?

One knows that one gluon exchange between quarks has a tensor term. The tensor term would induce a D-state admixture to 
the ground state of nucleon\cite{e}-\cite{i}. But this D-state admixture can not be more than a few percent at best. It definitely can not induce a D-state admixture of $25\%$. So where does this $25\%$ deformation of the nucleon in the ground state come from? As per our current understanding of confinement\cite{a}-\cite{c} the nucleon should be spherical. How come, it becomes so largely deformed? Seeking a mechanism for this within a consistent franework of QCD is an aim of this paper.

However if a colour singlet state exists it would mediate long range forces. From experiments we know that there are no 
such long range forces existing in nature. Hence this colour singlet state cannot be a representation of a gauge 
particle. However one may ask - could it manifest itself for physical reality in some other manner?

This problem may be resolved by assuming that this singlet state 
{\bf cannot act as a force propagator}
but it may manifest itself in some manner for describing say a static 
situation - such as for example in a bound 
system. In fact, as such it may provide a basis for confinement.
And indeed it does - as colour confinement for quark-antiquatrk mesonic states. This in fact is well known and well
accepted. {\bf Hence this colour singlet state finds application as a quantum mechanical state providing confinemet 
to a bound system ( like meson ) 
rather than being a representation of a mediator of a long range force.}

Let us write this in full

\begin{equation}
\sum_{i,j}{ \phi^a(i)\phi_a(j)} = \sum_{i \not= j}{ \phi^a(i)\phi_a(j)} + 
\sum_{i=j} {\phi^a(i)\phi_a(j)}
\end{equation}

where 'a' stands for colour and i,j stand for location of the colour 
fields in space. The first term provides the well 
known colour state for meson confinement. We shall show here as to how the 
second term provides the mising confinement 
term for baryons.

Let us look at colour singlet state for mesons
\begin{equation}
\phi^a(4)\phi_a(3)\Rightarrow\frac{1}{\sqrt{3}}\left(\bar{R}(4)R(3)+\bar{B}(4)B(3)
+\bar{G}(4)G(3)\right)
\end{equation}
Here we are explicitly exhibiting the locations of colour and anticolour. Here anticolour is situated at position 4 and colour at position 3. Reason for this somewhat odd numbering of positions will become clear below. Using the identity
\begin{equation}
\phi^a(4)=\epsilon^{abc}\phi_b(1)\phi_c(2)
\end{equation}
where the anticolour at position 4 creates colours at positions 1 and 2 which are in antisymmetric state. 
As we already know from the studies of diquark\cite{k} that position 4 should be the Centre of Mass of the positions 1 and 2 of identical quarks. Thus
\begin{equation}
\phi^a(4)\phi_a(3)=\epsilon^{abc}\phi_b(1)\phi_c(2)\phi_a(3)
\end{equation}
For baryons the positions 1, 2 ,3 at which the three colours are sitting there and which are all independent, give a totally antisymmetric state on the exchange of any two colours at positions 1, 2, and 3. These three independent positions 1, 2 and 3 are shown in Fig.1(a). All this is well understood for meson and baryon colour singletness.

Some further comments to explain what has been done so far. Mesons 
consisting of quark-antiquark are colour singlet as per eqn (3) above. 
Baryons consisting of 3 quarks are colour singlet due to the completely 
antisymmetric state given in eqn (5) above. Is there any connection 
between the colour singlet states of mesons (eqn (3)) and the colour 
singlet states of the baryons (eqn (5))?  As $3 \times 3 = \bar{3} + 6$
there indeed is! This connection is expressed by the unique  SU(3) 
transformation expessed in eqn (4) ( note that the same is not true in 
SU(2) ). This connection is well recognized by the picture of baryons as 
made up of a quark and a diquark. 
Diquark is nothing but what is given in 
eqn (4). This diquark picture is well studied and well rewieved in 
Ref [11] and for example
provides an explanation of the similarity of the Regge trajectories 
for mesons and baryons.

Next, we look at the colour singlet state given by the second term in eqn (2). 
Let us write this as ( labelling '3' for the arbitrary location ):
\begin{equation}
\frac{1}{\sqrt{3}}\left(\bar{R}(3)R(3)+\bar{B}(3)B(3)
+\bar{G}(3)G(3)\right)
\end{equation}
Now using Eq.(4) above
\begin{equation}
\frac{1}{\sqrt{6}}\left[\left(B(1)G(2)-G(1)B(2)\right)R(3)
+\left(G(1)R(2)-R(1)G(2)\right)B(3)
+\left(R(1)B(2)-B(1)R(2)\right)G(3)\right]
\end{equation}
where the anticolour at position 3 creates a two colour antisymmetric state at positions 1 and 2 respectively such that 
position 3 is the Centre of Mass of these. This is the point at which the third colour sits. This situation is depicted in 
Fig.1(b). Position 3 is rigid at the Center of Mass while position 1 and 2 can be exchanged. However in colour space, 
because of the Levi-Civita tensor $\epsilon^{abc}$ in Eq.(4) there is still antisymmetry between the exchange of any 
two colours. I will discuss this point below. Note that this state is colour singlet and has baryon number one as well. As per colour singlet hypothesis this colour singlet state for $B=1$ should contribute to confinement. Here we claim that this is the colour singlet confined state which has been missed so far for the confinement of baryons. 

If $r$ is the distance between particle 3 and particle 1 and 2, respectively (Fig.1(b)) then clearly orbital terms like $r^2$, $r^4$, $r^6.....$ or even powers of $r$ will occur. These corresponds to angular momentum states $L=2,4,6.....$ respectively. We note that for both $J=1/2$ and $J=3/2$ baryons these will provide states of Regge trajectory and thereby forming a natural explanation of the same in this model.
\begin{figure}[t]
\includegraphics[width=12cm]{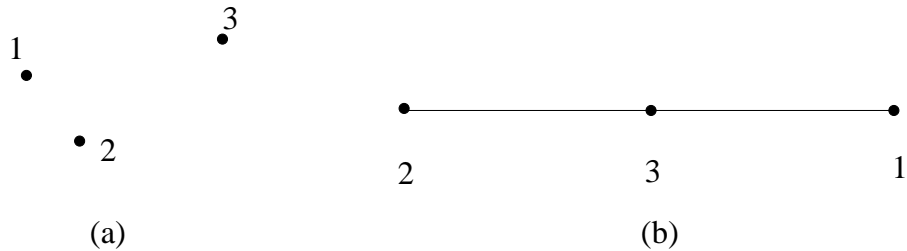}
\caption{Location of quarks in positions 1, 2 and 3 for (a) as per Eq.(4) and (b) as for Eq.(5) and Eq.(6)}
\end{figure}
Next note that $L=2$ with proper combination with the spin degrees of the three quarks will form the sought after D-state admixture as indicated above (see Appendix). Hence this gives a consistent mechanism for the D-state admixture of the nucleons in the ground state and shows that as per colour singlet confinement hypothesis, there are two states, one provides the S-state (already known) plus the new one giving the D-state admixture.

So we see that one colour singlet state has been missed for baryons. It is the new state and as discussed above, it is this which is making nucleon and delta and all the other baryons deformed. This new state also could explain the Regge trajectory states as well.

In Eq.(5) i.e. $\epsilon^{abc}\phi_b(1)\phi_c(2)\phi_c(3)$ is the well known colour antisymmetric state, the 
antisymmetry is imposed by the Levi-Civita term $\epsilon^{abc}$ where $a$, $b$ and $c$ are colour modes of $SU(3)_c$. 
However, the same antisymmetry is obtained if we do not change the colour states but exchange the position labels 1, 2 and 
3. The duality is due to the fact that these positions label quarks with all the properties i.e. flavour, spin, orbital 
state and colour state also. So colour sits on quarks and any exchange of one lead to the other. 

Next, the new colour singlet state is depicted in Eq.(6) and Eq.(7) can be seen in terms of Eq.(4) again provided the 
positions under 1 and 2 only being exchanged while 3 sits at the Center of Mass (Fig.1(b)). Full antisymmetry in the colour space arises, independent of the position at 3, due to the Levi-Civita term $\epsilon^{abc}$. Hence here notice that colour degrees of freedom break up from the quarks other degrees of freedom so as to provide antisymmetry.

This means that the new symmetry as imposed upon the system is
antisymmetric $SU(3)_C$ space and symmetric in the rest of the space. This symmetry is produced by having $1\leftrightarrow 2$ exchange only i.e. the mixed symmetric states like $\rho$- and $\lambda$- (see Appendix). Note that in Eq.(A.8) the $\rho -\rho$ and $\lambda -\lambda$ terms each are separately symmetric only in $1\leftrightarrow 2$ exchange. Only after being added do they provide symmetry with respect to '3'. And it is this that gives the D-state admixture to the baryon in the ground state. 

The subtleties of emerging colour (antisymmetry), spin and flavour (symmetry) symmetries can be understood in the following way. First in $SU(3)_C$ space write mixed colour state $\left[\phi_a(1)\phi_b(2)-\phi_b(1)\phi_a(2)\right]\phi_c(3)$. This is antisymmetric in exchange $a\leftrightarrow b$ and no specific symmetry with respect to colour 'C'. The same situation with respect to locations $1\leftrightarrow 2$ antisymmetry and none with respect to the exchange of 3. Now let the situation be that of Fig.1(b) when 3 is rigid and non movable Center of Mass of positions 1 and 2 which in t
urn can be exchanged.

Next insert the Levi-Civita antisymmetrization tensor 
$\epsilon^{abc}\left[\phi_a(1)\phi_b(2)-\phi_b(1)\phi_a(2)\right]\phi_c(3)$. Now this induces antisymmetry with respect to 
the exchange of colour 'C' also. As position 3 is rigid, the position '3' does not change but this exchange is in colour 
space.

Next in the quark flavour, spin and orbital space in Eq.(A-8)
\begin{eqnarray}
\chi^\lambda\left[\chi^S\psi^2_{2m\lambda}\right]^{J=1/2}=\frac{1}{\sqrt{2}}\left(\Psi_S+\Psi_\lambda\right)\\
\chi^\rho\left[\chi^S\psi^2_{2m\rho}\right]^{J=1/2}=\frac{1}{\sqrt{2}}\left(\Psi_S-\Psi_\lambda\right)
\end{eqnarray}
which have only symmetry with respect to $1\leftrightarrow 2$ and none with respect to flavour-spin at position 3. 
Clearly these are spurious excitations with respect to the Center of Mass in the Harmonic Oscillator Model. 
These spurious excitations are gotten rid of, and pure intrinsic excitations arise only on summing the above two as in 
Eq.(A-8). This induces new symmetry with respect to flavour-spin at the location 3. Hence what we show geometrically in 
Fig.1(b) is {\bf exactly} what is obtained when only intrinsic excitations with respect to the Center of Mass are 
calculated as in Eq. (A-8).

Note that for confinement we are getting two colour singlet states - one colour anti-symmetric in $SU(3)_C$ 
(giving spherically symmetric configuration of three quarks) and the other also colour anti-symmetric in $SU(3)_C$ 
(giving deformed 
D-state admixed state of three quarks). The requirement of colour singletness is unable to distinguish between the 
two - both being required at fundamental level. Therefore, just being singlet in colour can not be the only condition for 
confinement of quarks. Note that in the second case quarks are antisymmetric in colour space inspite of the fact that one 
of them is always rigidly located at the Center of Mass position with respect to the other two. The system still manages 
to give the corresponding totally symmetric state in the other degrees of freedom of the three quarks. The only thing 
common between the two colour singlet states is the requirement of colour antisymmetry in $SU(3)_C$. Just the colour 
singletness would suffice for the confinement of quark-antiquark in a meson. However, it appears that the more basic 
property for confinement of quarks in baryons is that of antisymmetry in the colour space. We need not demand of colour 
singletness as an additional requirement. Antisymmetry in colour space would itself ensure colour singletness for three 
quarks as done above for the two singlet states. 

So it may be summarized that one part of the colour singlet state was 
missed out when one only concentrates upon the canonical  
antisymmetric colour states for baryons. That gave a spherically symmetric nucleon. The missed color singlet state for 
baryon provides a D-state admixture to the baryons wave function and which makes the baryon deformed in the ground state 
itself. What we found on fundamental grounds here, matches and is consistent with the phenomenological quark model 
calculations of deformed baryons done earlier. 

\appendix*
\section{}
The nucleonic ground state as given in Eq.(1) is
\begin{equation}
|56,0^+\rangle\equiv|N_S\rangle
\end{equation}
where
\begin{equation}
|N_S\rangle=\frac{1}{\sqrt{2}}\left(\chi^\rho\phi^\rho+\chi^\lambda\phi^\lambda\right)\psi^0_{00S}
\end{equation}
where for example
\begin{equation}
\phi^\lambda_p=-\frac{1}{\sqrt{6}}\left(udu+duu-2uud\right)
\end{equation}
\begin{equation}
\phi^\rho_p=\frac{1}{\sqrt{2}}\left(udu-duu\right)
\end{equation}
and
\begin{equation}
\chi^\lambda_{\frac{1}{2}}=-\frac{1}{\sqrt{6}}\left(\uparrow\downarrow\uparrow+
\downarrow\uparrow\uparrow-2\uparrow\uparrow\downarrow\right)
\end{equation}
\begin{equation}
\chi^\rho_{\frac{1}{2}}=\frac{1}{\sqrt{2}}\left(\uparrow\downarrow\uparrow
-\downarrow\uparrow\uparrow\right)
\end{equation}
And
\begin{equation}
|70,2^+\rangle\equiv|N_D\rangle
\end{equation}
where
\begin{equation}
|N_D\rangle=\frac{1}{\sqrt{2}}\left\{\phi^\rho\left[\chi^S\psi^2_{2m\rho}\right]^{J=\frac{1}{2}}+
\phi^\lambda\left[\chi^S\psi^2_{2m\lambda}\right]^{J=\frac{1}{2}}\right\}
\end{equation}
where for example $(X=\rho \mbox{or} \lambda)$
\begin{equation}
\left[\chi^S\psi^2_{2mX}\right]^{J=\frac{1}{2}}_{J_Z=\frac{1}{2}}=
-\frac{1}{\sqrt{10}}\chi^S_{\frac{3}{2}}\psi^2_{2-1X}
+\frac{1}{\sqrt{5}}\chi^S_{\frac{1}{2}}\psi^2_{20X}
-\frac{3}{\sqrt{10}}\chi^S_{-\frac{1}{2}}\psi^2_{21X}
+\frac{2}{\sqrt{5}}\chi^S_{-\frac{3}{2}}\psi^2_{22X}
\end{equation}
For full details see Ref.\cite{c} and Ref.\cite{d}-\cite{i}.

\end{document}